\documentstyle[jkas]{article}

\runningauthor {Sung et al.} \year{2008} \volume{41}
\beginpage{1}\endpage{6}
\runningtitle{Coronal Temperature}

\begin{document}

\title{Coronal Temperature as an Age Indicator}

\author{Hwankyung Sung$^1$, M. S. Bessell$^2$, and Hugues Sana$^3$}
\address{$^1$ Department of Astronomy and Space Science, Sejong University,
	Kunja-dong 98, Kwangjin-gu, Seoul 143-747, Korea \\
	{\it e-mail}: sungh@sejong.ac.kr}

\address{$^2$ Research School of Astronomy and Astrophysics, Australian
	National University, MSO, Cotter Road, Weston, ACT 2611, Australia
	{\it e-mail}: bessell@mso.anu.edu.au}

\address{$^3$ European Southern Observatory, Casilla 19001, Santiago 19, Chile\\
	{\it e-mail}: hsana@eso.org}

\address{\normalsize{\it (Received October 23, 2007; Accepted January 15, 2008)}}
\offprints{H. Sung}

\abstract{
The X-ray spectra of late type stars can generally be well fitted by
a two temperature component model of the corona. We find that the temperatures
of both components are strong functions of stellar age, although the
temperature of the hotter plasma in the corona shows a larger scatter
and is probably affected by the activity of stars, such as flares.
We confirm the power-law decay of the temperature of the hot plasma, but the 
temperature of the cool plasma component decays linearly with $\log (age)$. }


\maketitle

\section{Introduction}

Since the launch of {\it ROSAT}, the X-ray properties of late type stars,
especially late type stars in open clusters, have been extensively studied
and are now well understood in terms of a stellar activity - rotation -
age paradigm. In most cases the investigators have derived
 the X-ray luminosities
and X-ray to bolometric luminosity ratios of stars, the X-ray luminosity
function of clusters, and compared them with other clusters (see Franciosini
et al. (2003) for example). Owing to the better light collecting power
and high angular resolution of the {\it Chandra} X-ray
Observatory and XMM-{\it Newton}, it is now possible to identify the X-ray
sources unambiguously and to determine the temperature of stellar coronae.

Previously it was thought that the coronal temperature depended on the
spectral-type of stars (Jordan \& Montesinos 1991) and showed a clear 
difference between that of F type stars and that of G/K type stars.
But the coronal temperatures of G and K type stars during quiescence are 
very similar (see Fig. 1, see also Fig. 5 of Pillitteri et al. 2006).

Schmitt et al. (1990) published an extensive analysis of the X-ray properties
of late-type stars based on data obtained with the {\it Einstein}
Observatory. They showed that models employing continuous emission measure
distributions provide equally adequate and physically meaningful and
more plausible descriptions of the phenomena.
Recently Marino et al. (2005) showed that the X-ray spectra of the G,
K and M type stars in IC 2391 are well described by two thermal components, 
although the instrument was capable of identifying up to 3 thermal components.

In this paper, we compile the coronal temperature of G and K (up to M,
if possible) type stars in open clusters for cases where the quality of the 
data was good enough to fit the X-ray spectra with a two temperature model 
and where, if possible, the X-ray spectra was not affected by a flaring event.
In this paper we highlight the possibility that coronal temperature is an age
indicator. In \S 2 and \S 3, we describe the data selection and derive the
relation between the average X-ray temperature and the age of open clusters.
A possible reason for the existence of such a relationship is
speculated in \S 4.

\section{Data}

\begin{figure}[t]
\epsfxsize=8.5cm \epsfbox{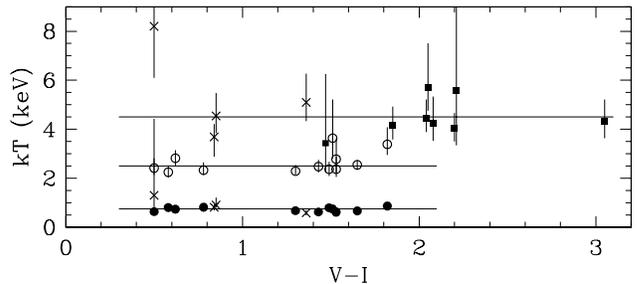}
\caption{Coronal temperatures of
late type stars in NGC 6231 (Sana et al.2007).
Filled and open circles represent, respectively, the lower and higher 
temperatures from two-temperature model fits of  spectra unaffected by 
flaring events, while crosses denote those affected by the flare. The 
squares are the temperatures of flares from a single temperature model fit.}
\end{figure}

\begin{table*}[t]
\begin{center}
\bf{\sc  Table 1.}
\sc{Average coronal temperatures of open clusters and stars}
\label{tbl-1}

\begin{tabular}{l@{}c@{}c@{}cccl}
\\ \hline \hline
Cluster & Age (Myr) & Observatory & $<kT_1 >$ & $<kT_2>$ & N &
source and criteria for data selection \\ \hline
Orion & $<$1.0$^{(1)}$ & CXO & 0.798 & 2.897 & 209 & Getman et al. (2005) : Sp $\geq$ G, \\
& & & $\pm$0.156 & $\pm$1.520 & & $\chi^2 \leq$ 1.50, d.o.f $\geq$ 21, no flare \\
NGC 2264 & 3.1$^{(2)}$ & CXO & 0.87 & 2.32 & 8 & Flaccomio et al. (2006) : F$_X \geq$ -13.0 \\
& & & $\pm$0.13 & $\pm$0.86 & & \\
NGC 6231 & 5.0$^{(3)}$ & XMM-{\it Newton} & 0.72 & 2.87 & 12 & Sana et al. (2007) : $V-I \geq 0.5$ \\
& & & $\pm$0.09 & $\pm$0.53 & & no flare \\
GJ 3305 & 13$^{(4)}$ & CXO & 0.6 & 2.8 & 1 & Feigelson et al. (2006) \\
NGC 2547 & 30$^{(5)}$ & XMM-{\it Newton} & 0.61 & 1.45 & 4 & Jeffries et al. (2006) \\
& & & $\pm$0.03 & $\pm$0.21 & & \\
IC 2391 & 53$^{(6)}$ & XMM-{\it Newton} & 0.40 & 1.10 & 4 & Marino et al. (2005) \\
& & & $\pm$0.09 & $\pm$0.11 & & \\
Pleiades & 115$^{(7)}$ & XMM-{\it Newton} & 0.44 & 1.07 & 11 & Briggs \& Pye (2003) \\
& & & $\pm$0.12 & $\pm$0.14 & & \\
NGC 2516 & 160$^{(8)}$ & XMM-{\it Newton} & 0.52 & 1.67 & 15 & Pillitteri et al. (2006) : d.o.f. $\geq$ 20 \\
& & & $\pm$0.18 & $\pm$0.74 & & \\
Praesepe & 650$^{(9)}$ & XMM-{\it Newton} & 0.40 & 0.90 & 2 & Franciosini et al. (2003) \\
& & & $\pm$0.01 & $\pm$0.01 & & \\
Sun & 4,600$^{(10)}$ & ASCA & 0.216 & 0.565 & - & Peres et al. (2000) : for solar maximum \\
& &(simulated) & $\pm$0.002 & $\pm$0.019 & & \\ \hline
\end{tabular}
\end{center}
Source of age -- $^{(1)}$ : Hillenbrand (1997), $^{(2)}$ : Sung, Bessell, \& Chun (2004),
$^{(3)}$ : Sung et al. (2008, in preparation), $^{(4)}$ : Feigelson et al. (2006),
$^{(5)}$ : Jeffries et al. (2006), $^{(6)}$ : Barrado y Navascu\'es et al. (1999),
$^{(7)}$ : Basri, Marcy, \& Graham (1996), $^{(8)}$ : Sung et al. (2002),
$^{(9)}$ : Franciosini et al. (2003), $^{(10)}$ : Barnes (2007)
\end{table*}

Getman et al. (2005) published results from extensive
observations of the Orion nebular cluster with the {\it Chandra}
X-ray Observatory. In their Tables 6 and 9, they presented the X-ray
properties of late-type stars in their quiescent state. We averaged
the coronal temperatures of 209 stars of spectral-types later than G, whose 
spectra could be fitted with two thermal components, and whose
reduced chi square ($\chi^2$) was less than 1.5 and degree of freedom
was larger than 21. As in Fig. 11 of Preibisch et al. (2005),
the lower temperature component shows no dependence on the stellar
photospheric temperature, although the higher temperature component
does show a weak dependence on photospheric temperature.
In the current analysis we disregarded the weak dependency on photospheric
temperature because we could find no such dependency in data from other
open clusters or stars.

For NGC 2264, we selected the X-ray temperatures of 8 stars whose
observed X-ray flux ($\log F_X \geq -13.0$) from Flaccomio et al. (2006).
If we lessen this criterion
to $\log F_X \leq -13.5$ (but still exclude the data whose higher
temperature is an upper limit or the error is an upper limit),
the average temperatures are slightly lower and so provide a better fit
to the relation in Fig. 2.

For NGC 6231, we selected the XMM-{\it Newton} data
from Sana et al. (2007). As shown in Fig. 1, we took an average for
the stars whose spectra were not affected by a flaring event.
For the nearby young star GJ 3305, Feigelson et al. (2006) presented
the coronal temperature without providing the errors.

For the Pleiades,
Daniel et al. (2002) presented the results from the {\it Chandra} X-ray
Observatory, but they assumed the temperature of the higher temperature
component was $kT_2$ = 3.5. We only took the Briggs \& Pye (2003) 
XMM-{\it Newton} data. Pillitteri et al. (2006) presented extensive
observations of NGC 2516 with XMM-{\it Newton}. We took the average temperature
of X-ray sources whose degree of freedom of spectral fit was greater than 20.

Franciosini et al. (2003) presented results from the spectral analysis of three
bright stars in Praesepe based on data obtained with XMM-{\it Newton}
observation. Two stars met our selection criteria. Stern et al. (1994) published
the coronal temperature of the stars in the Hyades. As the data were
obtained with {\it ROSAT}/ PSPC, the temperatures are systematically lower
and so the data were not taken into account.

Peres et al. (2000) simulated
the synthetic {\it ROSAT}/ PSPC and {\it ASCA}/SIS spectra of the sun
and presented the resulting temperatures. As {\it ROSAT}/ PSPC gives
systematically lower temperatures and {\it ASCA}/SIS has similar spectral
response to {\it Chandra} or XMM-{\it Newton}, we took the simulated
temperature from {\it ASCA}/SIS for the solar maximum.

\section{Coronal Temperature and Stellar Age}

\begin{figure}
\begin{center}
\epsfxsize=8.0cm \epsfbox{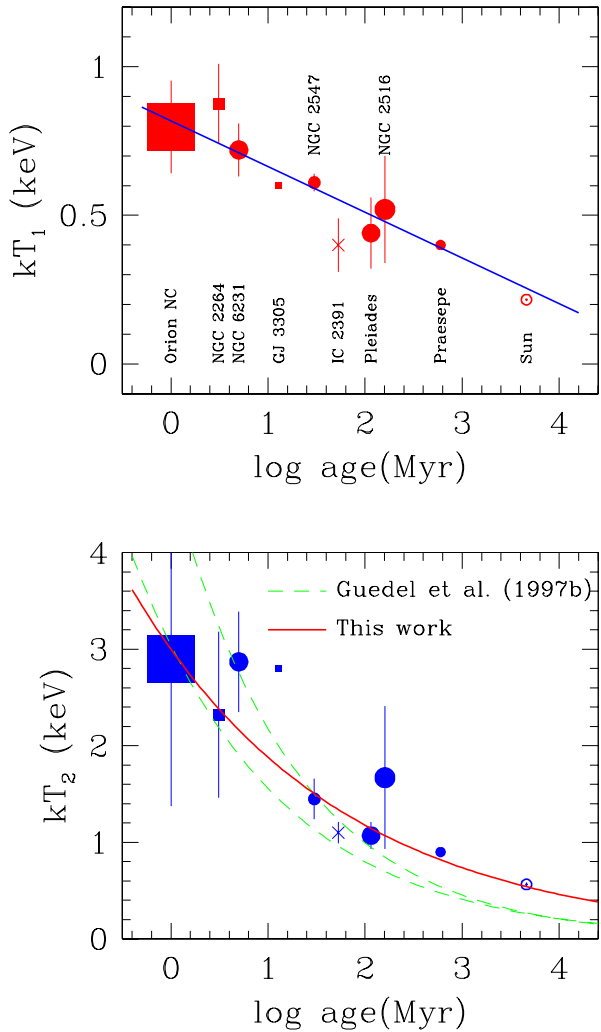}
\end{center}
\caption{Coronal temperatures of stars and the ages of open clusters.
(a) lower temperature versus age relation. (b) higher temperature versus
age relation. Squares and dots represent respectively the data obtained with
the {\it Chandra} X-ray Observatory and XMM-{\it Newton}, while the coronal
temperature of the sun is from simulated ASCA data for the solar maximum.
The cross represents the data for IC 2391 which were excluded in the regression.
The size of a symbol is proportional to the number of stars used in
calculating the average temperature. The solid and dashed lines represent
respectively the relations from this study and those from G\"udel et al.
(1997b). }
\end{figure}

Using the data compiled from the literature, we drew the relation
between age and coronal temperature shown in Fig. 2. The relation
between the average temperature of the lower temperature component and
the age of the cluster is quite good. Only the stars in IC 2391 show a
large deviation.

We fitted the following linear regressions to the data (excluding the data for
IC 2391).
In the regression we applied weights to the data points. The weights applied
were proportional to the square root of the number of stars used in the average.

\begin{equation}
kT_1 = 0.818 (\pm 0.025) - 0.154 (\pm 0.018) \cdot \log \tau({\rm Myr})
\end{equation}

\begin{equation}
\log( kT_2 ) = 0.477 (\pm 0.032) - 0.203 (\pm 0.025) \cdot \log \tau({\rm Myr})
\end{equation}

G\"udel et al. (1997b) derived a similar relation for the hotter component
using the {\it ROSAT} data for solar-type stars. But the power they obtained
was -0.34 (for a Raymond-Smith model) or -0.29 (for a MEKAL model) (see the 
dashed lines in Fig. 2), rather than -0.203 ($\pm$ 0.025) in equation (2). 
They used the data for older stars (the youngest star among them is EK Dra, 
a Pleiades moving group member- about 100 Myr). In addition
the temperature from {\it ROSAT} is rather lower than that determined using
the data obtained with {\it Chandra} or XMM-{\it Newton}. As can be seen
in Fig. 2, the slopes for $\tau$ (age) $\geq$ 100 Myr are very similar.

The value of both the low and high temperature components and the
age of open clusters are seen to be well correlated, but the relation for
the higher temperature component shows a somewhat larger scatter.
As can be seen in Fig. 1, the lower
temperature component is nearly independent of flaring events.
Although the X-ray spectra are affected by the flaring events, the temperature
of the cool component of the plasma is very similar to that of the other stars
whose spectra are not affected by a flaring event. On the other hand,
the temperature of the hot component is strongly affected by the flare
(see for the case of YY Gem : G\"udel et al. (2001b) or AB Dor :
G\"udel et al. (2001a)).

Among the clusters and stars in Table 1, NGC 6231, 2516, and
a binary system GJ 3305 have higher values of kT$_2$.
Sung et al. (2002) suggested that the high X-ray activity among the stars
in NGC 2516 may be related to the high binary frequency of the cluster.
Close binary systems such as RS CVn type binaries show strong,
long-lasting X-ray activity owing to the tidal forces they exert on each other.
The large scatter in the temperature of the hot plasma could
therefore be interpreted either as resulting from flaring activity or from
the binarity.

\section{Discussion}

\subsection{IC 2391}

The data for IC 2391 show a large deviation from the regression line
and is neglected in the regression in \S 3. The four X-ray emission
stars in IC 2391 are slightly brighter at a given ($B-V$) or ($V-I$).
Their location in the color-magnitude diagram is also strongly affected by
contamination due to background stars in the Perseus arm (see Fig. 2 
of Patten \& Pavlovsky (1999) or Fig. 2 of Rolleston \& Byrne 1997).
In addition, the X-ray temperature of these late type stars is
lower than that of the early type stars (VXR 46 \& 56 in IC 2391)
whose X-ray emission is due to the coronal emission from late type companions.
In the Pleiades,
two early type stars (HII 1234 \& 1384 - Briggs \& Pye 2003) show similar
temperatures (kT$_1$ \& kT$_2$) to those of late type stars.

It is interesting that the coronal temperature of VXR 37 is very low even 
though the spectrum was affected by a flare. As seen in Fig. 1,
the temperature of flares of stars in NGC 6231 is much higher, and even
the temperature of a solar-like flare is between kT = 1.2 -- 4.4 keV
(Kobayashi et al. 2003; Peres et al. 2000). We suggest that either these
stars are not members of IC 2391 or their temperatures have been determined 
incorrectly.

\subsection{Possible Interpretation}

G\"udel et al. (1997a) analyzed the X-ray and extreme ultraviolet spectra of
the young solar analog EK Draconis and found that the distribution
of the differential emission measure was essentially bimodal. The two peaks
they found was interpreted as simply reflecting the variation with energy
of the radiative cooling function of a thermal, optically thin plasma.
The emission measure distribution of AB Dor (Sanz-Forcada, Maggio \& Micela
2003) also shows two prominent peaks at $\log T_e$ = 6.9 and 7.3.
In fact, Gehrels \& Williams (1993) suggested that the bimodal distribution of
coronal temperature of late type stars could be caused by the two local
positive slopes in the cooling curve of an optically thin plasma.
But our data shown in Fig. 2 do not show any clustering near
0.6 keV or near 0.1 keV. The change in the average coronal temperature
is rather a monotonic function of cluster age. And therefore the phenomena
we found cannot be interpreted as resulting from the radiative cooling
curve of a thin plasma.

Barnes (2003a) extensively studied the rotational evolution of
solar- and late-type stars and identified the existence of two
sequences in the rotational period vesus color plane. He confirmed
the ``so-called'' Skumanich style spin-down process for late-type stars
having a dominant Sun-like, {\it interface} magnetic field. In addition
he identified a {\it convective} sequence of fast rotators in open clusters
that possessed only a convective field, which was not only unable to
deplete angular momentum but also incapable of coupling the surface
convection zone to the inner radiative zone. Barnes (2003b) attempted
to interpret the relation between X-ray emission and rotation in terms of
his classification of rotating stars.

%
G\"udel et al. (1997b) successfully reproduced the hot temperature tail of
the coronal differential emission measure using the time evolution of the
physical properties of solar flares. From Fig. 1, we could also surmise that
the hot plasma component is associated with flaring events.
But the flares involved with the hot temperature plasma may comprise
many successive microflares that cannot be resolved with
point-source observations.

On the other hand, the decline in the temperature of the cool component cannot
be explained by the usual flaring events. Parker (1988) suggested nanoflares
as a possible heating source for the solar X-ray corona.
Another proposed mechanism of heating the solar corona is heating by
Alfv\'en waves due to the inhomogeneity of the magnetic field (Heyvaerts \&
Priest 1983).
Although this mechanism cannot explain the occurrence of flares, it could
heat a relatively wide region. As can be seen in Fig. 2, although
both processes have a dependency on the strength of the magnetic field
and consequently could explain the time evolution of coronal temperature,
the physical process involved in the heating of the cool plasma could differ
from that of the hot plasma. 

Very recently De Pontieu et al. (2007) found (at least) two types of spicules
in the solar limb from the time-series Ca II H filter observation with the
Solar Optical Telescope on board
the Japanese solar mission {\it Hinode}. While type-I spicules are less dynamic
and show in many cases a parabolic motion, type-II spicules are very dynamic,
have a shorter lifetime and show apparently fast upward speeds between
40 $km~ s^{-1}$ and 300 $km~ s^{-1}$. Many spicules of both types undergo
significant transverse motions which can be interpreted as Alfv\'enic motions.
They suggested that the Alfv\'en waves carry an energy flux and play
a significant role in the heating of the corona of the quiet Sun and in 
accelerating the solar wind. Although we have no definite grounds for 
rejecting nanoflares as a possible heating source of the cool component, we 
favour Alfv\'en waves as the source of heating of the cool plasma component
of the stellar coronae.

\section{Conclusions}

From this study, we find that

(1) The temperature of the cool component of the coronal plasma is less affected
by stellar activity and is a better indicator of stellar age. The coronal
temperature of the cool component decreases linearly with the $\log (age)$ of
a star/cluster.

(2) The temperature of the hot component can also be used as an indicator of
stellar age, but is affected by stellar activity, such as flares.

(3) The X-ray temperature of the hot plasma decreases with about the -0.2
power of stellar age. This slope from our data is
greater than that obtained by G\"udel et al. (1997b) (about -0.3).

(4) The difference in the age-dependency of the coronal temperature of the
hot and cool plasmas could indicate differences in the heating mechanisms of
the two components. We favour Alfv\'en waves as the source of heating of the 
cool plasma. 

\acknowledgments

H. S. acknowledges the support of the Korea Science and Engineering
Foundation to the Astrophysical Research Center for the Structure and
Evolution of the Cosmos (ARCSEC$''$) at Sejong University. H.S. would like
to express his thanks to J. C. Chae at Seoul National University and
D. H. Lee at Kyung Hee University for valuable discussion.

\end{document}